\documentclass[aps,prl,superscriptaddress,showpacs,floatfix,twocolumn]{revtex4}

\usepackage{graphicx}	

\begin{document}

\title{Nuclear Modification Factors for Hadrons At Forward and Backward
Rapidities in Deuteron-Gold Collisions at $\sqrt{s_{NN}}$ = 200\,GeV}

\newcommand{\abilene}{Abilene Christian University, Abilene, TX 79699, USA}
\newcommand{\acadsin}{Institute of Physics, Academia Sinica, Taipei 11529, Taiwan}
\newcommand{\banaras}{Department of Physics, Banaras Hindu University, Varanasi 221005, India}
\newcommand{\barc}{Bhabha Atomic Research Centre, Bombay 400 085, India}
\newcommand{\bnl}{Brookhaven National Laboratory, Upton, NY 11973-5000, USA}
\newcommand{\caucr}{University of California - Riverside, Riverside, CA 92521, USA}
\newcommand{\ciae}{China Institute of Atomic Energy (CIAE), Beijing, People's Republic of China}
\newcommand{\cns}{Center for Nuclear Study, Graduate School of Science, University of Tokyo, 7-3-1 Hongo, Bunkyo, Tokyo 113-0033, Japan}
\newcommand{\colorado}{University of Colorado, Boulder, CO 80309}
\newcommand{\columbia}{Columbia University, New York, NY 10027 and Nevis Laboratories, Irvington, NY 10533, USA}
\newcommand{\dapnia}{Dapnia, CEA Saclay, F-91191, Gif-sur-Yvette, France}
\newcommand{\debrecen}{Debrecen University, H-4010 Debrecen, Egyetem t{\'e}r 1, Hungary}
\newcommand{\elte}{ELTE, E{\"o}tv{\"o}s Lor{\'a}nd University, H - 1117 Budapest, P{\'a}zm{\'a}ny P. s. 1/A, Hungary}
\newcommand{\fsu}{Florida State University, Tallahassee, FL 32306, USA}
\newcommand{\gsu}{Georgia State University, Atlanta, GA 30303, USA}
\newcommand{\hiroshima}{Hiroshima University, Kagamiyama, Higashi-Hiroshima 739-8526, Japan}
\newcommand{\ihepprot}{Institute for High Energy Physics (IHEP), Protvino, Russia}
\newcommand{\illuiuc}{University of Illinois at Urbana-Champaign, Urbana, IL 61801}
\newcommand{\isu}{Iowa State University, Ames, IA 50011, USA}
\newcommand{\jinrdubna}{Joint Institute for Nuclear Research, 141980 Dubna, Moscow Region, Russia}
\newcommand{\kek}{KEK, High Energy Accelerator Research Organization, Tsukuba-shi, Ibaraki-ken 305-0801, Japan}
\newcommand{\kfki}{KFKI Research Institute for Particle and Nuclear Physics (RMKI), H-1525 Budapest 114, POBox 49, Hungary}
\newcommand{\korea}{Korea University, Seoul, 136-701, Korea}
\newcommand{\kurchatov}{Russian Research Center ``Kurchatov Institute", Moscow, Russia}
\newcommand{\kyoto}{Kyoto University, Kyoto 606, Japan}
\newcommand{\labllr}{Laboratoire Leprince-Ringuet, Ecole Polytechnique, CNRS-IN2P3, Route de Saclay, F-91128, Palaiseau, France}
\newcommand{\lawllnl}{Lawrence Livermore National Laboratory, Livermore, CA 94550, USA}
\newcommand{\losalamos}{Los Alamos National Laboratory, Los Alamos, NM 87545, USA}
\newcommand{\lpc}{LPC, Universit{\'e} Blaise Pascal, CNRS-IN2P3, Clermont-Fd, 63177 Aubiere Cedex, France}
\newcommand{\lund}{Department of Physics, Lund University, Box 118, SE-221 00 Lund, Sweden}
\newcommand{\muenster}{Institut f\"ur Kernphysik, University of Muenster, D-48149 Muenster, Germany}
\newcommand{\myongji}{Myongji University, Yongin, Kyonggido 449-728, Korea}
\newcommand{\nagasaki}{Nagasaki Institute of Applied Science, Nagasaki-shi, Nagasaki 851-0193, Japan}
\newcommand{\newmex}{University of New Mexico, Albuquerque, NM 87131, USA }
\newcommand{\nmsu}{New Mexico State University, Las Cruces, NM 88003, USA}
\newcommand{\ornl}{Oak Ridge National Laboratory, Oak Ridge, TN 37831, USA}
\newcommand{\orsay}{IPN-Orsay, Universite Paris Sud, CNRS-IN2P3, BP1, F-91406, Orsay, France}
\newcommand{\peking}{Peking University, Beijing, People's Republic of China}
\newcommand{\pnpi}{PNPI, Petersburg Nuclear Physics Institute, Gatchina, Russia}
\newcommand{\riken}{RIKEN (The Institute of Physical and Chemical Research), Wako, Saitama 351-0198, JAPAN}
\newcommand{\rikjrbrc}{RIKEN BNL Research Center, Brookhaven National Laboratory, Upton, NY 11973-5000, USA}
\newcommand{\saopaulo}{Universidade de S{\~a}o Paulo, Instituto de F\'{\i}sica, Caixa Postal 66318, S{\~a}o Paulo CEP05315-970, Brazil}
\newcommand{\seoulnat}{System Electronics Laboratory, Seoul National University, Seoul, South Korea}
\newcommand{\stonybrkc}{Chemistry Department, Stony Brook University, Stony Brook, SUNY, NY 11794-3400, USA}
\newcommand{\stonycrkp}{Department of Physics and Astronomy, Stony Brook University, SUNY, Stony Brook, NY 11794, USA}
\newcommand{\subatech}{SUBATECH (Ecole des Mines de Nantes, CNRS-IN2P3, Universit{\'e} de Nantes) BP 20722 - 44307, Nantes, France}
\newcommand{\tenn}{University of Tennessee, Knoxville, TN 37996, USA}
\newcommand{\titech}{Department of Physics, Tokyo Institute of Technology, Tokyo, 152-8551, Japan}
\newcommand{\tsukuba}{Institute of Physics, University of Tsukuba, Tsukuba, Ibaraki 305, Japan}
\newcommand{\vandy}{Vanderbilt University, Nashville, TN 37235, USA}
\newcommand{\waseda}{Waseda University, Advanced Research Institute for Science and Engineering, 17 Kikui-cho, Shinjuku-ku, Tokyo 162-0044, Japan}
\newcommand{\weizmann}{Weizmann Institute, Rehovot 76100, Israel}
\newcommand{\yonsei}{Yonsei University, IPAP, Seoul 120-749, Korea}
\newcommand{\deceased}{\dagger}
\affiliation{\abilene}
\affiliation{\acadsin}
\affiliation{\banaras}
\affiliation{\barc}
\affiliation{\bnl}
\affiliation{\caucr}
\affiliation{\ciae}
\affiliation{\cns}
\affiliation{\colorado}
\affiliation{\columbia}
\affiliation{\dapnia}
\affiliation{\debrecen}
\affiliation{\elte}
\affiliation{\fsu}
\affiliation{\gsu}
\affiliation{\hiroshima}
\affiliation{\ihepprot}
\affiliation{\illuiuc}
\affiliation{\isu}
\affiliation{\jinrdubna}
\affiliation{\kek}
\affiliation{\kfki}
\affiliation{\korea}
\affiliation{\kurchatov}
\affiliation{\kyoto}
\affiliation{\labllr}
\affiliation{\lawllnl}
\affiliation{\losalamos}
\affiliation{\lpc}
\affiliation{\lund}
\affiliation{\muenster}
\affiliation{\myongji}
\affiliation{\nagasaki}
\affiliation{\newmex}
\affiliation{\nmsu}
\affiliation{\ornl}
\affiliation{\orsay}
\affiliation{\peking}
\affiliation{\pnpi}
\affiliation{\riken}
\affiliation{\rikjrbrc}
\affiliation{\saopaulo}
\affiliation{\seoulnat}
\affiliation{\stonybrkc}
\affiliation{\stonycrkp}
\affiliation{\subatech}
\affiliation{\tenn}
\affiliation{\titech}
\affiliation{\tsukuba}
\affiliation{\vandy}
\affiliation{\waseda}
\affiliation{\weizmann}
\affiliation{\yonsei}
\author{S.S.~Adler}	\affiliation{\bnl}
\author{S.~Afanasiev}	\affiliation{\jinrdubna}
\author{C.~Aidala}	\affiliation{\columbia}
\author{N.N.~Ajitanand}	\affiliation{\stonybrkc}
\author{Y.~Akiba}	\affiliation{\kek} \affiliation{\riken}
\author{A.~Al-Jamel}	\affiliation{\nmsu}
\author{J.~Alexander}	\affiliation{\stonybrkc}
\author{K.~Aoki}	\affiliation{\kyoto}
\author{L.~Aphecetche}	\affiliation{\subatech}
\author{R.~Armendariz}	\affiliation{\nmsu}
% ...
\collaboration{PHENIX Collaboration} \noaffiliation

\date{\today}

\begin{abstract}

We report on charged hadron production in deuteron-gold reactions at
$\sqrt{s_{NN}}=200$~GeV.  Our measurements in the deuteron-direction cover
$1.4 < \eta < 2.2$, referred to as forward rapidity, and in the
gold-direction $-2.0 < \eta < -1.4$, referred to as backward rapidity, and
a transverse momentum range $p_{T}=0.5-4.0$~GeV/$c$. We compare the
relative yields for different deuteron-gold collision centrality classes.
We observe a suppression relative to binary collision scaling at forward
rapidity, sensitive to low momentum fraction ($x$) partons in the gold
nucleus, and an enhancement at backward rapidity, sensitive to high
momentum fraction partons in the gold nucleus.
\end{abstract}



\maketitle



Deep inelastic scattering of leptons on the proton revealed the
proton's substructure of point-like parton
constituents~\cite{bjork69}.  This substructure, usually
described quantitatively as Parton Distribution Functions,
evolves as one probes the proton at shorter wavelength or
equivalently higher momentum transfer, $Q^2$. 
Using the measured quark and antiquark distribution functions and the 
DGLAP~\cite{DGLAP} and BFKL~\cite{BFKL} evolution equations, a strong 
increase in the gluon density is expected at high $Q^{2}$ and small $x$ (fraction
of the proton momentum carried by the parton).
Such an increase is indeed observed at HERA~\cite{HERA_ALL}, suggesting that
at sufficiently small $x$, gluons should overlap in space and time.
This overlap should result in gluon fusion, and thus reduce the gluon
density at low $x$ and enhance it at larger $x$.  This gluon fusion limits the
achievable gluon density, leading to gluon saturation.  This saturation is sometimes
described as the formation of a Color Glass Condensate (CGC)~\cite{CGC_FIRST}.  Gluon
saturation is expected to be a larger effect in nuclei where the partons
from different nucleons overlap as well.  Suppression of low $x$ partons in nuclei
relative to nucleons has been experimentally observed and is referred to as
nuclear shadowing~\cite{shadowing}.  However, this shadowing is often described in terms of
modification of the leading-twist parton densities in nuclei~\cite{twist}.

In 2003 the Relativistic Heavy Ion Collider (RHIC) collided deuteron
and gold nuclei at $\sqrt{s_{NN}} = 200$\,GeV. At this energy,
most hadrons with $p_T > 2.0$\,GeV/$c$ arise
from parton-parton interactions and can be used as a probe of nuclear
partonic structure.  
Hadrons with $p_T > 2.0$\,GeV/$c$ at forward rapidity $1.4 < \eta < 2.2$ 
are sensitive to low $x$ partons in the gold nucleus $0.001 < x < 0.03$.
Hadrons at backward rapidity $-2.0 < \eta < -1.4$ are sensitive to high $x$ 
partons in the gold nucleus $0.04 < x < 0.5$.
It has been predicted that gluon saturation at small $x$
will suppress hadronic yields at forward rapidity~\cite{CGC_second} 
with the transverse momentum scale for the
onset of the gluon saturation set by $Q_s^2 [{\rm GeV}^2] =
0.13\;N_{coll}e^{\lambda y}$~\cite{CGC_third} for $d+Au$ collisions at
RHIC. Here $\lambda \sim 0.3$ is determined from HERA
data~\cite{COLOR_DIPOLE} and $N_{coll}$ is the number of
nucleon-nucleon inelastic collisions.  Thus, for central collisions
and within our forward rapidity coverage $Q_s^2$ is expected
to be of order $2-4~{\rm GeV}^2$ and may have observable consequences.
Novel hadron production mechanisms, such as quark recombination~\cite{hwa},
can also impact the distribution of particles in the forward rapidity region.

Results on charged hadron yields at forward
rapidity from the BRAHMS experiment have shown a suppression of the yield of hadrons
in central, compared to peripheral, $d+Au$ collisions~\cite{brahms_rcp}.
At mid-rapidity, PHENIX has reported a modest enhancement of the yield of hadrons with $p_{T} > 1.5$\,GeV/$c$~\cite{phenix_nosup}.  
This enhancement, generally referred to as the
``Cronin effect'' is often ascribed to initial state scattering of the parton traversing the nucleus
prior to the high $Q^{2}$ scattering~\cite{cronin}.  At backward rapidities (large x), 
anti-shadowing and other effects of the surrounding nuclear medium (e.g. the EMC effect)~\cite{Review_NME} 
may compete, making predictions challenging.

It is important to note that in the transverse momentum range of this measurement, 
$0.5<p_T<4.0$\,GeV/$c$, hadron production is also sensitive to
soft physics phenomena which are determined by coherent hadron-hadron
interactions.
In $p+A$ reactions at lower energies soft hadron production shifts from 
forward to backward rapidity, with a larger shift for larger nuclear targets. 
Thus, at low $p_{T}$ one may observe an increase (decrease) in hadron yields at backward (forward)
rapidity which is not necessarily a reflection of changes at the partonic structure level.

In this Letter, we present results from the PHENIX experiment~\cite{phenixnim} on the
ratio of hadron yields at forward and backward pseudorapidity 
for different centrality classes of $d+Au$ collisions. 
PHENIX has two spectrometers designed for measuring muon production
over the pseudorapidity range $-2.2 < \eta <-1.2$ (backward spectrometer) 
and $1.2 < \eta <2.4$ (forward spectrometer)~\cite{phenixnim}.
The spectrometers start with a thick hadron absorber
comprised of 19~cm of brass and 60~cm of low-carbon steel
between the collision point and active
detectors along the beam axis, primarily to reduce hadronic background
for muon measurements~\cite{PHENIX_MAGNET}.
After this material, the Muon Tracker (MuTr) detector, consisting of three stations of cathode strip
chambers, tracks charged particles in a magnetic field.  
The momentum
resolution is 5\% (for typical momenta in this analysis) and the absolute scale is known to better than 1\%.
Following the muon magnet backplate (30~(20)~cm of steel in the forward (backward) spectrometer) 
there is a Muon Identifier (MuID) detector.
The MuID consists of five layers of planar drift tubes interleaved with layers of
steel for further hadron absorption (10~cm thick in the first two layers and 20~cm thick for the
remaining layers).  The layers are numbered 0-4, with 4 being the most downstream.
The MuID is used to separate muons from hadrons and provide triggering capabilities.

Although these spectrometers were designed to detect muons, they can
also be used to measure charged hadrons via two independent methods. 
The first method
is via the identification of hadrons which penetrate part way through the 
MuID, referred to as ``punch-through hadrons.''  The second method is via muons
from light mesons $\pi,K$ which decay before interacting in the
absorber material. By measuring these decay muons, we can reconstruct
the yield of their parent light mesons.  
In both of these methods the absolute yield of hadrons is difficult to determine
due to uncertainties in the punch-through and decay probabilities.
However, this small probability is independent of $d+Au$ collision centrality, and
thus not knowing the absolute yields does not affect the precision 
of measured ratios of hadron yields between the different
classes of events.

\begin{figure}[tbh]
\includegraphics[width=1.0\linewidth]{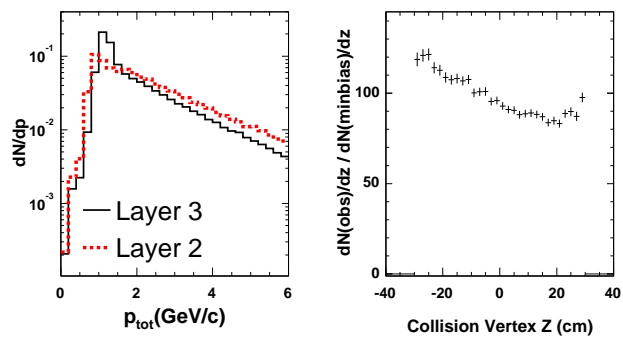}
\caption{\label{fig:muon_vtx} (color online).
 (left) The total momentum $p_{tot}$ measured in the MuTr without 
energy loss
correction of all charged tracks penetrating to MuID layers 2 (grey) and 3 (black).
(right) Collision vertex distribution for events with muons at forward rapidity,
corrected for the minimum bias collision vertex distribution.}
\end{figure}

The punch-through hadron identification is achieved by 
studying particles that stop somewhere within the MuID before the last
layer.
For muons penetrating up to layer 2 and 3 we expect the average momenta measured in the MuTr 
of $p = 1.0$\,GeV/$c$ and $p = 1.2$\,GeV/$c$, respectively,
corresponding to the average ionization energy loss in traversing the spectrometer
material.  
The reconstructed momentum distributions for particles stopping in layers
2 and 3 of the MuID are shown in Figure~\ref{fig:muon_vtx}.  In addition to the expected
muon peaks, there is a broad distribution extending to higher momentum which
is the result of punch-through hadrons.  These hadrons also suffer ionization
energy loss up to the relevant layer and then suffer an inelastic collision
in the MuID steel and do not penetrate further.
We thus select a clean sample of hadrons by demanding that a
track stop in MuID layer 2 or 3 and have momentum more than
$3\sigma$ away from the muon peaks.  
Muon contamination in our sample is estimated from simulations to be less than $5\%$.  

Another source of background for the punch-through hadrons is 
secondary particles produced from hadronic showering
in the absorber. This background is reduced by requiring that the track point back
to the primary collision vertex, as determined from the Beam-Beam Counter (BBC).
The background from secondary particles varies as a function of $p_T$ and 
is typically $\sim 1-5\%$ of the signal based on simulations.
We also apply acceptance cuts $-2.0 < \eta <-1.4$ and $1.4 < \eta <2.2$
in order to reduce background at small angles.

Some hadrons will decay into muons before the absorber,
and the decay muons are then measured by the muon spectrometers.
Muons can result from many sources including decays
of $\pi$, $K$, $D$ mesons, and $J/\psi$.  These particles have
a finite decay probability $P_{decay}$ before they reach the absorber
\begin{equation}
 P_{decay}(p,L) = 1 - e^{-\frac{L \cdot m}{\tau \cdot p}}
\end{equation}
where $L \sim 41$cm, is the distance from the collision vertex to the absorber; 
$p$, $m$ and $\tau$ are the momentum, mass and proper lifetime of the parent particle. 

Thus, collisions that occur far from the absorber will be more likely to produce
muons from light meson decays than those that occur close to the absorber. Charm
hadrons, however, due to their very short proper decay lengths, 
$e^{-\frac{L \cdot m}{\tau \cdot p}} \ll 1$, will have minimal collision vertex
dependence . The right panel of Figure~\ref{fig:muon_vtx} shows the collision vertex
distribution from events in which muons are detected at forward rapidity, corrected
for the minimum bias collision vertex distribution.  The large vertex dependence
indicates a significant fraction of the muons are from pion and kaon decay.  Using
this distribution, we can separate the muons from pion and kaon decay from other
contributions.  
The acceptance and efficiency vary by less than 5\% over the z vertex range, which
establishes an upper bound on the systematic error attributable to the subtraction
of these non-signal contributions.
It should be noted that the measured muon $p_{T}$ is approximately
15\% lower on average than the parent hadron $p_{T}$, which is not corrected for in this analysis.

The data set for this analysis was collected under two different trigger conditions.  
We recorded $67 \times 10^{6}$ minimum bias triggers which required at least one hit
in both the PHENIX forward $3.0 < \eta < 3.9$ and backward $-3.9 < \eta < -3.0$
Beam-Beam Counters (BBC) and a reconstructed vertex position within $|z|<30$ cm
along the beam axis.  The minimum bias trigger accepts $88 \pm 4\%$ of all inelastic
$d+Au$ collisions~\cite{phenix_nosup}. The second data set, sampling $5.3 \times
10^{9}$ minimum bias events, was collected with the MuID trigger which requires at
least one track penetrating the first four layers of the MuID.

We divide these events into four centrality classes based on the number of particle
hits in the backward BBC counter covering $-3.9 < \eta < -3.0$. Using a Glauber
model~\cite{phenix_nosup} and simulation of the BBC, we determine the average number
of binary collisions in each centrality class. The classes are categorized as
follows: $60-88\%$ ($\langle N_{coll} \rangle =3.1 \pm 0.3$), $40-60\%$ ($\langle
N_{coll} \rangle =7.0 \pm 0.6$), $20-40\%$ ($\langle N_{coll} \rangle =10.6 \pm
0.7$),and $0-20\%$ ($\langle N_{coll} \rangle=15.4 \pm 1.0$).


There is a correlation between having a particular physics process (for example
the production of a high $p_{T}$ hadron) and the BBC response.  The BBC coverage in pseudorapidity is well
separated from the muon spectrometers so the correlation is not predominantly due to jet
fragmentation, but rather an underlying event correlation.  We have studied this effect
in detail using proton-proton and $d+Au$ data, and in simulations, and have accounted for this
correlation bias.  The bias correction factors we apply range from 0-7\% depending on the centrality
category and the physics process.  The systematic errors on these corrections are less
than 4\%.

\begin{figure}[tbh]
\includegraphics[width=1.0\linewidth]{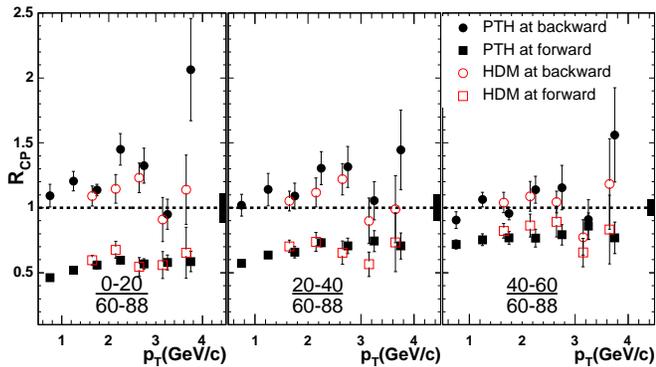}  
\caption{\label{fig:rcp_pt} (color online).
$R_{cp}$ as a function of $p_T$ at forward rapidity (squares)
and backward rapidity (circles) for different centrality classes.
}
\end{figure}

\begin{figure}[tbh]
\includegraphics[width=1.0\linewidth]{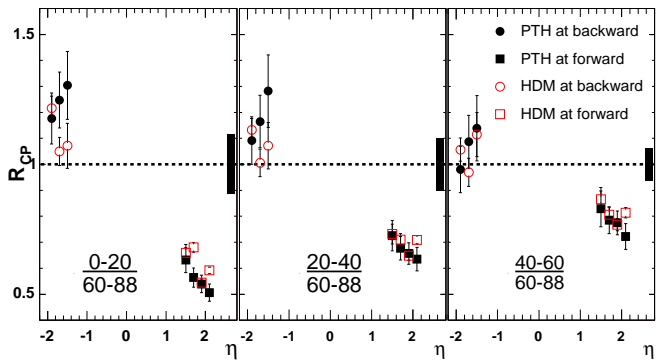}  
\caption{\label{fig:rcp_eta} (color online).
$R_{cp}$ as a function of $\eta$ for $1.5 < p_T < 4.0$\,GeV/$c$ for different centrality classes. 
}
\end{figure}

The {\it nuclear modification factor} $R_{cp}$ is defined as the ratio
of the particle yield in central collisions to the particle yield in
peripheral collisions, each normalized by the average number of nucleon-nucleon
binary collisions ($\langle N_{coll} \rangle$):
\begin{equation}
  R_{cp} = \frac{ \langle \left( \frac {dN}{d\eta dp_T} \right)^{Central} \rangle / \langle N_{coll}^{Central} \rangle }  {\langle \left(\frac{dN}{d\eta dp_T}\right)^{Peripheral} \rangle / \langle N_{coll}^{Peripheral} \rangle }
\label{eq:rcp}
\end{equation}
The hadron $R_{cp}$, using the most peripheral centrality class ($60-88\%$) for
normalization, is shown in Figure~\ref{fig:rcp_pt} as a function of $p_T$ at forward
and backward rapidities. The results from the punch-through hadron (PTH) and hadron
decay muon (HDM) techniques are both shown and are in quite good agreement. We also
show the results integrated over $1.5 < p_T < 4.0$\,GeV/$c$ as a function of
pseudorapidity in Figure~\ref{fig:rcp_eta}.

There are two types of systematic uncertainties in our analysis. Common systematic
errors which move all data points up and down together include the error on
$\frac{N^{central}_{coll}}{N^{periperal}_{coll}}$ (10.8\% for the most central bin),
the centrality bias correction factors (4\%), and the centrality-dependent tracking
efficiency (4\%) determined by embedding Monte Carlo particles in real data.  
Common systematic errors are shown as a black bar. Point-to-point systematic errors
result from sensitivities to analysis cuts and are $5-10\%$.  They are added in
quadrature with the statistical errors and shown as error bars.

It is notable that our two measurement methods have different sensitivity to
different hadrons.  The particle composition ($\pi/K/p$ ratio) of the observed sample is
modified relative to the particle composition at the collision vertex due to
species-dependent nuclear interaction cross-sections affecting the punch-through
hadrons and due to species-dependent decay lifetimes affecting the hadron decay
muons.  Both effects enhance the kaon contribution to our $R_{cp}$ measurements.  
The uncertainty on our charged hadron $R_{cp}$ values introduced by this effect is
estimated to be less than $4\%$ by calculating the difference between the kaon
$R_{cp}$ and inclusive charged particle $R_{cp}$ determined by PHENIX at
mid-rapidity~\cite{PHENIX_PID_dAu}.

We observe that $R_{cp}$ shows a suppression at forward rapidity that is largest for
the most central events.  The opposite trend is observed at backward rapidity where
$R_{cp}$ shows an enhancement that is also largest for the most central events.  We
observe a weak $p_{T}$ dependence with slightly smaller $R_{cp}$ values at lower
$p_{T}$. We observe a clear pseudorapidity dependence at forward rapidity with
$R_{cp}$ dropping further at larger $\eta$ values.  Within our current uncertainties
we are unable to discern any pseudorapidity dependence at backward rapidity.

In Figure 4 we compare results from the BRAHMS experiment~\cite{brahms_rcp} with our
results at forward rapidity.  The PHENIX data and the BRAHMS data are in agreement
within systematic uncertainties.

\begin{figure}[tbh]
\includegraphics[width=1.0\linewidth]{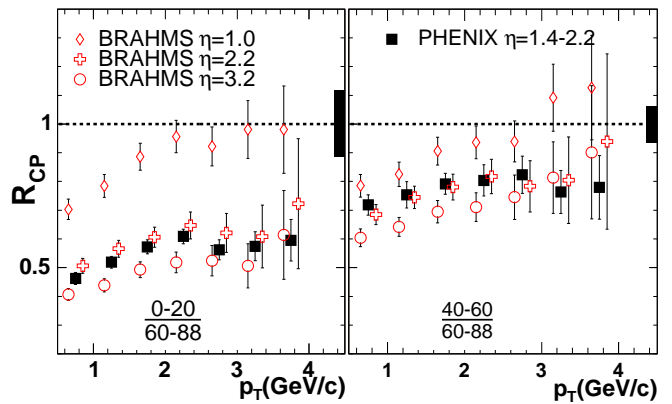}  
\caption{\label{fig:rcp_brahms} (color online).
PHENIX $R_{cp}$ as a function of $p_{T}$ at forward rapidities shown as the average
of the two methods.  Note that the BRAHMS results are for negative hadrons at
$\eta=2.2,3.2$ and their centrality ranges ($0-20\%/60-80\%$ and $30-50\%/60-80\%$)
are somewhat different from ours. 
} 
\end{figure}
 
The suppression of hadron yields relative to binary collision scaling at forward
rapidity is expected from initial state nuclear effects.  However, detailed
comparisons with various theoretical approaches is necessary in order to
discriminate between different models.  In particular the lack of a strong $p_{T}$
depedence at both forward and backward rapidities must be understood as the physics
processes transition from ``soft'' to ``hard'' physics scales.


To summarize, we observe a suppression in hadron yields relative to binary collision
scaling at forward rapidities and an enhancement at backward rapidity for central
relative to peripheral $d+Au$ reactions at $\sqrt{s_{NN}} = 200$\,GeV. The forward
rapidity suppression is in qualitative agreement with the expectation of shadowing
and saturation effects in the small $x$ region in the gold nucleus.  However, other
physics effects must also be considered in understanding the full $p_{T}$ and $\eta$
dependence. The source of the backward rapidity enhancement, and the possible
contribution of anti-shadowing of large $x$ partons, has yet to be understood.


We thank the staff of the Collider-Accelerator and Physics
Departments at BNL for their vital contributions.  We acknowledge
support from the Department of Energy and NSF (U.S.A.), 
MEXT and JSPS (Japan), CNPq and FAPESP (Brazil), NSFC (China), 
IN2P3/CNRS, CEA, and ARMINES (France), 
BMBF, DAAD, and AvH (Germany), 
OTKA (Hungary), DAE and DST (India), ISF (Israel), 
KRF and CHEP (Korea), RMIST, RAS, and RMAE (Russia), 
VR and KAW (Sweden), U.S. CRDF for the FSU, 
US-Hungarian NSF-OTKA-MTA, and US-Israel BSF.


\def\Journal#1#2#3#4{{#1}{\bf #2}, #3 (#4)}
\def\IJMPA{{Int. J. Mod. Phys.}~{\bf A}}
\def\JPG{{J. Phys}~{\bf G}}
\def\NCA{Nuovo Cimento\ }
\def\NIM{Nucl. Instrum. Methods\ }
\def\NIMA{{Nucl. Instrum. Methods\ }~{\bf A}}
\def\NPA{{Nucl. Phys.}~{\bf A}}
\def\NPB{{Nucl. Phys.}~{\bf B}}
\def\PLB{{Phys. Lett.}~{\bf B}}
\def\PLC{Phys. Repts.\ }
\def\PR{Phys. Rev.\ }
\def\PRL{Phys. Rev. Lett.\ }
\def\PRD{Phys. Rev. D\ }
\def\PRC{Phys. Rev. C\ }
\def\RMP{Rev. Mod. Phys.\ }
\def\SPJ{Sov. Phys. JETP\ }
\def\SJNP{Sov. J. Nucl. Phys.\ }
\def\ZPC{{Z. Phys.}~{\bf C}}

\end{document}